\documentclass{article}
\usepackage{spconf,amsmath,amsfonts,graphicx,bm,siunitx,color,fancyhdr}

\newcommand{\x}{\bm{x}}
\newcommand{\z}{\bm{z}}
\newcommand{\h}{\bm{h}}
\newcommand{\vc}{\bm{c}}
\newcommand{\C}{\bm{C}}
\newcommand{\F}{\bm{F}}
\newcommand{\RR}{\mathbb{R}}
\newcommand{\CC}{\mathbb{C}}
\renewcommand{\L}{\mathcal L}
\newcommand{\vmu}{\bm{\mu}}
\newcommand{\veps}{\bm{\epsilon}}
\newcommand{\vsig}{\bm\sigma}
\newcommand{\vdel}{\bm{\delta}}
\newcommand{\diff}{\mathrm d}
\newcommand{\qphi}{{q_{\phi}(\z|\x)}}
\newcommand{\pth}{{p_{\theta}(\x|\z)}}
\newcommand{\xz}{\bm{x}|\bm z}
\newcommand{\zx}{\bm{z}|\bm x}
\newcommand{\id}{\mathbf{I}}

\DeclareMathOperator{\E}{E}
\DeclareMathOperator{\KL}{D_{KL}}
\DeclareMathOperator{\diag}{diag}
\DeclareMathOperator*{\tran}{^{\mkern-1.5mu{T}}}
\DeclareMathOperator*{\herm}{^{\mkern-1.5mu{H}}}

\title{CSI CLUSTERING WITH VARIATIONAL AUTOENCODING}
\name{Michael Baur\thanks{This work was supported by the Bavarian Ministry of Economic Affairs, Regional Development and Energy within the project 6G Future Lab Bavaria.}, Michael W\"urth, Michael Koller, Vlad-Costin Andrei, Wolfgang Utschick}
\address{Department of Electrical and Computer Engineering, Technical University of Munich, Germany\\
Email: \{mi.baur, michael.wuerth, michael.koller, vlad.andrei, utschick\}@tum.de}

\begin{document}
\thispagestyle{fancy}
\maketitle
\begin{abstract}
The model order of a wireless channel plays an important role for a variety of applications in communications engineering, e.g., it represents the number of resolvable incident wavefronts with non-negligible power incident from a transmitter to a receiver. Areas such as direction of arrival estimation leverage the model order to analyze the multipath components of channel state information. In this work, we propose to use a variational autoencoder to group unlabeled channel state information with respect to the model order in the variational autoencoder latent space in an unsupervised manner. We validate our approach with simulated 3GPP channel data. Our results suggest that, in order to learn an appropriate clustering, it is crucial to use a more flexible likelihood model for the variational autoencoder decoder than it is usually the case in standard applications.
\end{abstract}
%


\section{Introduction}
\label{sec:intro}

Wireless channels are characterized by different multipath components, each comprising individual parameters such as directions of arrival (DoA) at the receiver~\cite{Akdeniz2014}. Every impinging wavefront at a receiver antenna array from a transmitter represents one of the multiple propagation paths. The number of resolvable propagation paths is typically related to the model order and both play an important role in many fields of communications engineering such as DoA estimation~\cite{Krim1996} or channel estimation with compressive sensing~\cite{Berger2010}, where the model order or the number of resolvable paths is often assumed to be known. In this work, we investigate the use of generative modeling with variational autoencoders (VAE) for the clustering of unlabeled channel state information~(CSI) with respect to its number of resolvable paths.

The idea of clustering wireless channels is an ongoing topic of research. An early work of multipath clustering is performed in~\cite{Yu2004} by first employing a generalized expectation-maximization (EM) algorithm, which delivers the clusters. Second, the obtained clusters are determined by inspection. In~\cite{He2017a}, a Kernel-power-density based algorithm is proposed to solve this clustering problem. The algorithm allows for clustering of the multipath components, e.g. path delay or DoA. Another framework adapts the variational Bayesian EM algorithm towards the multipath clustering task~\cite{Li2020}. All of the above methods have in common that clustering is performed on the basis of the (extracted) multipath components, but not with respect to the number of resolvable paths and not directly on realizations of the vector channels. 

The task of clustering data, in our case wireless channels, can be interpreted as grouping data that are similar and separating data that are dissimilar with respect to a predefined criterion. Representation learning based on unsupervised machine learning (ML) is a candidate to fulfill this objective as it aims to find a meaningful representation of the data that facilitates the search for beneficial information~\cite{Bengio2013}. ML has already been widely applied in communications engineering and will be important for the development of future wireless systems. 
The VAE~\cite{Kingma2014} is a suitable tool to generate artificial data from unknown distributions solely represented by training samples. A particular feature of the VAE is the manifestation of a structured representation of the underlying unlabeled data by means of clustering in the latent space. 
These clusters can be interpreted as representatives of different features of the data and can be used to generate new data with controllable features by the VAE decoder. The clustering capability of VAEs has been demonstrated for multiple image datasets. In this work, we transfer this result to CSI data and demonstrate the capability of the VAE for CSI clustering.

The contribution of this work is as follows. We show that a VAE can  cluster CSI data with respect to the number of paths, which we use interchangeably for the model order in this work, although knowing that non-resolvable paths collapse the determinable model order. To this end, we propose a VAE decoder that models a conditional Gaussian posterior with a circular covariance matrix compared to the standard choice which assumes a Gaussian posterior with a scaled identity covariance matrix. We observe that this choice of a more expressive VAE decoder significantly improves the ability of the VAE encoder to produce a better clustered output. In contrast to other methods, our method solely works with realizations of the vector channels and does not require any further preprocessing steps. 


\section{Channel Model}
\label{sec:system}

A base station with a uniform linear array (ULA) equipped with $M$ antennas is considered, which receives signals from a single antenna mobile terminal. Similar to~\cite{Neumann2018}, we assume conditionally Gaussian channels, which are frequency-flat and fast-fading. The channels are distributed according to
\begin{equation}
    \h \mid \vdel \sim \mathcal{N}_{\CC}(\vmu_{\vdel}, \C_{\vdel}), \qquad \vdel \sim p(\vdel)
    \label{eq:3gpp}
\end{equation}
with mean vector $\vmu_{\vdel} \in \CC^M$ and covariance matrix $\C_{\vdel} \in \CC^{M\times M}$. The random variable $\vdel$ with distribution $p(\vdel)$ describes some abstract prior which can be interpreted, for example, as the multipath components of the channel. For the duration of one coherence interval, the covariance matrix is assumed to be constant. Conditionally Gaussian channels are a common modeling choice in communication scenarios. They appear for example in the stochastic channel models of the 3rd Generation Partnership Project (3GPP)~\cite{3GPP2020}. In particular, the covariance matrices in the 3GPP channel models follow the relation $ \C_{\vdel} = \int_{-\pi}^{\pi} g(\theta;\vdel) \bm{a}(\theta) \bm{a}(\theta)\herm \diff\theta $, where $g(\theta;\vdel) \geq 0$ denotes a power angular spectrum and $\bm{a}(\theta)$ the steering vector corresponding to the array geometry.

It is known that a covariance matrix $\C_{\vdel}$ of a ULA with half-wavelength spacing exhibits a Toeplitz structure, which can be well approximated by a circular matrix for a large number of antennas~\cite{Neumann2018}. An interesting property of a circular matrix is that it is diagonalizable by the discrete Fourier transform (DFT) matrix $\F \in \CC^{M\times M}$, such that
$ \C_{\vdel} = \F\herm \diag(\vc_{\vdel}) \F $, $\vc_{\vdel} \in \RR^M $. Consequently, using DFT, the covariance matrix of the transformed channels $ \x = \F\h $ is diagonal and corresponds to $\diag(\vc_{\vdel})$. In Section~\ref{sec:vae}, we use the circular approximation of the Toeplitz covariance matrix of a ULA in conjunction with Fourier-transformed channel data to design a VAE which is suitable for the task of CSI clustering with respect to the number of resolvable paths.

\section{VARIATIONAL AUTOENCODER MODEL}
\label{sec:vae}

This section reviews the VAE after highlighting some important aspects of variational inference (VI). Subsequently, we present a modification of the VAE framework which enables to learn the circular covariance matrices of the conditional Gaussian channel model defined in Section~\ref{sec:system}.

\subsection{Variational Inference Background}
\label{subsec:vi}
The VAE has its origin in the field of VI~\cite{Blei2017}. The central term in VI is the evidence lower bound (ELBO) 
\begin{equation}
    \L(q) = \log p(\x) - \KL(q(\z) \| p(\z|\x)) \leq \log p(\x)
    \label{eq:elbo1}
\end{equation}
which involves the data log-likelihood $\log p(\x)$ and the Kullback-Leibler (KL) divergence between the variational distribution $q(\z)$ and the posterior $p(\z|\x)$,
that is, \\ $ \KL(q(\z) \| p(\z|\x)) = $ $\int q(\z) \log \frac{q(\z)}{p(\z|\x)} \diff\z$. 
The variational distribution $q(\z) \in \mathcal Q$ stems from a family of distributions $\mathcal Q$ and is supposed to match the posterior $p(\z|\x)$ as closely as possible. In the case of equivalence, the KL divergence vanishes and the ELBO is equal to the data log-likelihood. The main problem for finding a suitable $q(\z)$ is that the encountered integrals are usually intractable. VI circumvents this problem by turning the task of solving an integral into an optimization problem which reads as
\begin{equation}
    \min_{q(\z)\in\mathcal Q}\, \KL(q(\z) \| p(\z|\x)).
\end{equation}
A maximization of the ELBO therefore achieves two goals: the data log-likelihood is maximized and the best approximation in $\mathcal Q$ to $p(\z|\x)$ is found.

\subsection{VAE with Scaled Identity Covariance Matrix}
\label{subsec:scaledid}

\begin{figure}[t]
\begin{minipage}[b]{1.0\linewidth}
  \centering
  \centerline{\includegraphics{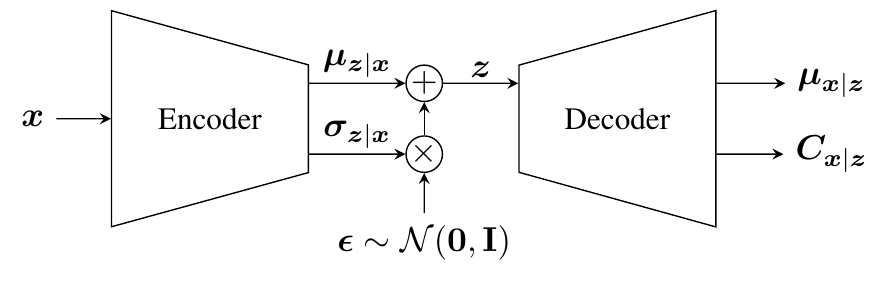}}
  \label{fig:umap_data}
\end{minipage}
\vspace*{-7.5mm}
\caption{Sketch of a VAE with Gaussian posteriors.}
\label{fig:vae}
\vspace*{-5mm}
\end{figure}

The VAE, independently proposed in~\cite{Kingma2014} and~\cite{Rezende2014}, makes VI accessible to a broad audience. The reason for this is that a VAE consists of a relatively simple structure and is scalable to large data. This is achieved by using the reparameterization trick and neural networks to learn the parameters of the encountered distributions. In the context of a VAE, the variational distribution is usually explicitly conditioned on the data~$\x$, which yields $q(\z)=\qphi$. For the training of a VAE, the ELBO is often expressed as 
\begin{equation}
    \L(q) = \E_\qphi[\log \pth + \log p(\z) - \log \qphi]
    \label{eq:elbo2}
\end{equation}
where $\qphi$ represents the encoder network and $\pth$ the decoder network with neural network parameters $\phi$ and $\theta$, respectively~\cite{Kingma2019}. These networks are central components of a VAE and are shown in~Fig.~\ref{fig:vae}, which we explain in more detail later in this section. Since calculating the gradient with respect to the neural network parameters is difficult, the reparameterization trick is employed. It assumes that $\z$, which is also referred to as latent variable, is the result of a differentiable transformation of a random variable $\veps$, which belongs to a simpler distribution, e.g., $\veps \sim \mathcal{N}(\bm 0, \id)$. In~\cite{Kingma2019}, it is proposed to optimize \eqref{eq:elbo2} with a single noise sample $\veps$.

If we assume $\mathcal{N}_{\CC}(\vmu_{\xz}, \C_{\xz})$, $\mathcal{N}(\vmu_{\zx}, \diag(\vsig_{\zx}^2))$, and $\mathcal{N}(\bm{0}, \id)$, for $\pth$, $\qphi$, and $p(\z)$, respectively, and omit constant terms we can reformulate the ELBO as
\begin{multline}
    \L(q) = -\log\det\C_{\xz} - (\x-\vmu_{\xz})\herm \C_{\xz}^{-1}(\x-\vmu_{\xz}) \\
    + \frac{1}{2}\lVert\veps\rVert^2 + \bm{1}\tran\log\vsig_{\zx} - \frac{1}{2}\lVert\z\rVert^2
    \label{eq:elbogauss}
\end{multline}
with the all ones vector $\bm 1$. A sketch of a VAE with the previous assumptions is depicted in Fig.~\ref{fig:vae}. A data point is passed through the encoder to receive the vectors $\vmu_{\zx}$ and $\vsig_{\zx}$, which are in turn used to reparameterize $\z$. The resulting latent vector $\z$ is passed through the decoder to obtain the vector $\vmu_{\xz}$ and the matrix $\C_{\xz}$. These quantities could then be used to generate new samples $\tilde{\x} \sim \mathcal{N}_{\CC}(\vmu_{\xz}, \C_{\xz})$. More generally, a newly generated sample is supposed to follow the true distribution of the data $p(\x)$.

Since the size of outputs of the decoder for $\C_{\xz}$ scales with $M^2$ and the inverse $\C_{\xz}^{-1}$ in \eqref{eq:elbogauss} is expensive to compute, it is uncommon to work with full covariance matrices. Instead, it is often assumed that the covariance matrix is a scaled identity, i.e., $\C_{\xz} = \sigma_{\xz}^2\id$. It is obvious that this approach simplifies the computation of \eqref{eq:elbogauss} immensely. Its drawback is that the expressiveness of $\pth$ is limited.

\subsection{VAE with Diagonal Covariance Matrix}
\label{subsec:diag}
Only few methods exist in the literature to model the covariance matrix $\C_{\xz}$ more expressively since the focus is mainly on the optimization of the mean $\vmu_{\xz}$ instead of the whole distribution $\pth$.
One method uses a Cholesky decomposition of $\C_{\xz}$~\cite{Kingma2019}. Another rarely employed approach uses a diagonal covariance matrix $\C_{\xz}=\diag(\vc_{\xz})$, as  described in~\cite{Kingma2014}. It is worth investigating because it increases the expressiveness of $\pth$ and keeps the computation of \eqref{eq:elbogauss} efficient. With this assumption, we can further simplify \eqref{eq:elbogauss} to
\begin{multline}
    \L(q) = \bm{1}\tran \left( -\log\vc_{\xz} - \vc_{\xz}^{-1} \odot \lvert\x-\vmu_{\xz}\rvert^2 \right) \\
    + \frac{1}{2}\lVert\veps\rVert^2 + \bm{1}\tran\log\vsig_{\zx} - \frac{1}{2}\lVert\z\rVert^2.
    \label{eq:elbodiag}
\end{multline}
The vector $\vc_{\xz}^{-1}$ represents the elementwise reciprocals of the corresponding vector and $\odot$ represents an elementwise multiplication. At this point, we can draw the connection to the channel model presented in Section~\ref{sec:system}. We would like to learn a complete covariance matrix conditioned on every input channel. Since such a covariance matrix is too computationally expensive to learn, we propose to use its circular approximation in combination with DFT-transformed channels $\x$. This allows us to learn the diagonal covariance matrices of the transformed channels by training a VAE that maximizes Equation~\eqref{eq:elbodiag}. Note that a sample $\x$ is modeled as conditionally Gaussian with the VAE in Fig.~\ref{fig:vae}, with mean value $\vmu_{\xz}$ and diagonal covariance matrix $\C_{\xz}=\diag(\vc_{\xz})$. Also note that for the chosen dataset of conditionally Gaussian 3GPP channels in DFT domain, without loss of generality, the choice of a diagonal covariance matrix at the decoder output of the VAE exactly resembles the structure of the true covariance matrix. This relation further indicates why the proposed model is beneficial for our application as we can approximate the true distribution of our channel data much better than with a scaled identity covariance matrix.

\section{RESULTS}
\label{sec:results}

In this section, we first give an overview about implementation details. Subsequently, we present the channel clustering results with respect to the number of resolvable paths.

\subsection{Dataset Description and Network Architecture}
\label{subsec:dataset}

We create a dataset of wireless channels based on the elaborations in Section~\ref{sec:system} for the 3GPP urban-macro channel model. To this end, we randomly sample 33\,000 channels per model order, each belonging to a different covariance matrix. The number of antennas at the base station is set to 32. The first 30\,000 channels are used for training and the remaining 3\,000 channels for evaluation purposes. As this work is intended to be a first proof of concept, we create a dataset consisting only of channels with one or five propagation paths. The paths are generated without regard to their resolvability and with different path gains from a uniform distribution in the range of $[0,1]$ such that the individual path gains add up to one. This leads to the dataset comparison results shown in Table~\ref{tab:mmd}. The applied maximum mean discrepancy (MMD) is a kernel based measure for the dissimilarity of two datasets. It can be fully evaluated using samples taken from the respective distributions. For more details we refer the reader to~\cite{Gretton2012}. Based on the MMD, a hypothesis test can be designed to determine the dissimilarity of two datasets~\cite{Utschick2021}, which is displayed for channels with different numbers of paths in Table~\ref{tab:mmd}. Every table entry represents the true positive rate (TPR) that channels of the corresponding model orders belong to a different dataset. Clearly, within the considered margins, channels with only one path are most dissimilar to channels with five paths, which is the reason for the constitution of our dataset. The investigation of other cases is subject of future work.

\begin{table}[t]
    \centering
    \caption{MMD hypothesis test TPRs for a comparison of channel data exhibiting the respective number of paths.}
    \smallskip
    \begin{tabular}{c|ccccc}
          & 1 path & 2 paths & 3 paths & 4 paths & 5 paths \\ \hline
	  1 path  & 0.02 & 0.26 & 0.65 & 0.92 & 0.98 \\
	  2 paths & - & 0.01 & 0.14 & 0.40 & 0.60 \\
	  3 paths & - & - & 0.01 & 0.10 & 0.23 \\
	  4 paths & - & - & - & 0.02 & 0.06 \\
	  5 paths & - & - & - & - & 0.00
    \end{tabular}
    \label{tab:mmd}
\vspace*{-5mm}
\end{table}

The encoder and decoder are implemented with a symmetric structure. It consists of three layers, where each layer is built of a 1D convolutional layer with kernel size 7 and stride 2, followed by a batch normalization layer and ReLU activation function. In the encoder, the convolutional layers map from 1 to 8 to 32 to 128 convolutional channels. In the decoder, this happens vice versa. In both networks, this composition is followed by a fully connected layer to map to the correct dimension. The size of the latent space is set to~4. Here, we choose the lowest value for which we can still see distinct clusters in our experiments. Real and imaginary part of the complex valued channels are stacked to a vector, sized two times the amount of antennas. We use the Adam optimizer~\cite{Kingma2015} with a learning rate of \num{5.3e-4} and the method of free bits for optimization~\cite{Kingma2019}.

\subsection{Learned CSI Clusterings with VAE}
\label{subsec:cluster}

\begin{figure}[t!]

\begin{minipage}[b]{1.0\linewidth}
  \centering
  \centerline{\includegraphics{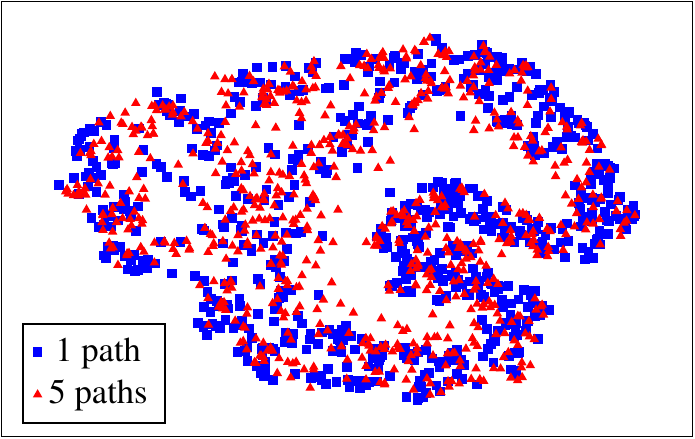}}
  \centerline{(a) UMAP of raw data.}\medskip
\end{minipage}

\begin{minipage}[b]{1.0\linewidth}
  \centering
  \centerline{\includegraphics{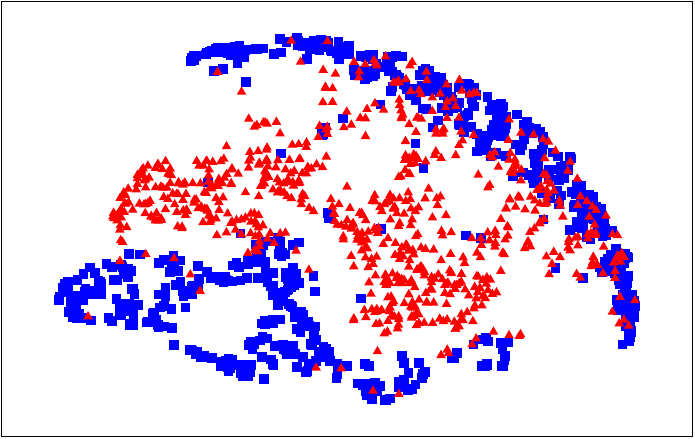}}
  \centerline{(b) UMAP of latent space of VAE with $\C_{\xz} = \sigma_{\xz}^2\id$.}\medskip
\end{minipage}

\begin{minipage}[b]{1.0\linewidth}
  \centering
  \centerline{\includegraphics{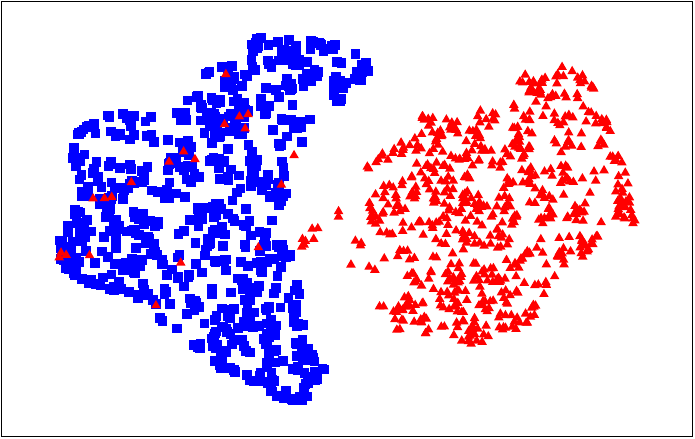}}
  \centerline{(c) UMAP of latent space of VAE with $\C_{\xz}=\diag(\vc_{\xz})$.}
\end{minipage}

\caption{UMAP embeddings for different types of data.}
\label{fig:umap_main}
\vspace*{-5mm}
\end{figure}

We show different 2D uniform manifold approximation and projection (UMAP) embeddings in this section~\cite{McInnes2018}. It should be noted at this point that UMAP embeddings look different depending on the chosen hyperparameters. Distances between and sizes of clusters are not representative. UMAP embeddings are supposed to preserve local and global structures. As they are not descriptive for the embeddings, we omit the axes in the following illustrations. Note that latent space plots were created using previously unseen evaluation data and that data labels were used solely to illustrate results but were not available during unsupervised learning of the VAE. 

We plot the UMAP embedding of the input channel data in Fig.~\ref{fig:umap_main}(a) as a comparison to see the benefit of the learned VAE clustering. Clearly, no clusters are observable in the raw data. 
In contrast, Fig.~\ref{fig:umap_main}(b) shows the UMAP embedding of the four-dimensional mean latent vectors $\vmu_{\zx}$ corresponding to each input data point. We trained a VAE with scaled identity covariance matrix at the output of the decoder for this embedding, as explained in Section~\ref{subsec:scaledid}. The VAE manages more poorly than not to separate the channels that belong to one path from the channels that belong to five paths. While the model orders are somehow separated, the learned VAE encoder embedding is not useful for clustering. 

When we used the non-standard VAE with diagonal covariance matrix at the output of the decoder on the same data, as highlighted in Section~\ref{subsec:diag}, the embedding changes dramatically. The channels with one path and five paths are clearly separated in the latent space in Fig.~\ref{fig:umap_main}(c). Only a few channels with five paths are projected onto the one path cluster. The most likely reason for these "outliers" is that the corresponding channels have one dominant path and four weak paths, so the channels are more alike to one-path channels. The remaining channels are distinctly assigned to their correct clusters. 

In Fig.~\ref{fig:umap_main}, we see that adjusting the decoder's implicit likelihood model affects the structure of the VAE latent space. 
By employing a diagonal covariance matrix model at the decoder output, the stochastic model becomes more flexible. This indirectly forces the encoder to produce a more structured latent space to exploit the improved flexibility of the decoder.

\section{CONCLUSION}
\label{sec:conclusion}

We have shown in this work that it is possible to cluster CSI data with respect to their model order, represented by the number of propagation paths, using unsupervised learning techniques by training a VAE. It turned out that the choice of the implicit likelihood model represented by the decoder is beneficial for clustering in the latent space of the VAE. We achieved this by subjecting the channel data based on a ULA to a DFT matrix, as well as by using a VAE that provides conditional diagonal covariance matrices at the output of the decoder. In our future work, we will address further open questions, e.g., the investigation of different channel models, the extension to MIMO channels, and the model order analysis of measured data from real-world measurement campaigns. Another topic will be the focus on the actual generation of channel data with controllable features using autoencoding; initial results for single antenna channel impulse responses in \cite{Weisser2021} point the way.

\vfill\pagebreak
\bibliographystyle{IEEEbib}
\bibliography{main}

\end{document}